# Distinctive Electronic Characteristics and Ultra-high Thermoelectric Power Factor in Be-Fe Intermetallics

Qi D. Hao[1, 2], H. Wang[2], Xiang R. Chen[1]*, Hua Y. Geng[2, 3]*

[1] *College of Physics, Sichuan University, Chengdu 610065, P. R. China;*

[2] *National Key Laboratory of Shock Wave and Detonation Physics, Institute of Fluid Physics, China Academy of Engineering Physics, Mianyang, Sichuan 621900, P. R. China;*

3 *HEDPS, Center for Applied Physics and Technology, and College of Engineering, Peking University, Beijing 100871, P. R. China;*

**Abstract**

Beryllium (Be) alloys are indispensable in cutting-edge applications due to their unique advantages. However, the scientific understanding about their structure and property is deficient, which greatly restricts their applications within a narrow field. In this work, a systematic investigation on the structure and properties of Be-Fe binary was carried out with first-principles unbiased evolutionary algorithms. Five new intermetallics unreported before, including insulating $Be_{11}Fe$ and $Be_4Fe$, metallic $Be_3Fe$, and metastable BeFe and $BeFe_2$ were discovered, among which $Be_{11}Fe$ has a unique clathrate structure and is an electride. Surprisingly, we found that Fe unexpectedly acts as an anion in all known Be-Fe intermetallics, and its valence state can even reach -5, leading to the complete filling of its 3$d$ orbitals. Most of these compounds exhibiting a gap or pseudogap at the Fermi level. Specifically, the band gap is determined as 0.22 eV and 0.85 eV for $Be_{11}Fe$ and $Be_4Fe$ at the level of single-shot GW, respectively. This is the first report of insulating phases in Be-based intermetallics. We also discovered that $Be_{11}Fe$ exhibits an impressive thermoelectric power factor of 178 $\mu W\ cm^{-1}\ K^2$ at room temperature, to our best knowledge, the highest among known semiconductors under ambient conditions, indicating its potential for waste heat harvesting and active cooling. These findings will deepen our understanding of Be-based and Fe-based compounds, and expand the application fields of Be-based alloys to a brand-new realm.

\* *Corresponding authors. E-mail: s102genghy@caep.cn; xrchen@scu.edu.cn*

**Introduction**

Beryllium (Be) has important applications in cutting-edge areas, especially in the aerospace industry and deep space exploration(*1, 2*), due to its superior mechanical property and performance. It possesses the highest stiffness, highest strength and melting temperature among all light metals, together with high specific heat, good thermal conductivity and a small thermal expansion coefficient(*3-6*). Despite its inherent toxicity, these exceptional properties have motivated efforts to overcome the manufacturing and usage challenges associated with Be. To facilitate its applications, alloys or composite such as Be-Cu(*7*), Be-Al(*8, 9*), Be-Ni(*10*), and Be-Ti(*11*) etc., have been developed. Nonetheless, all intermetallic compounds or alloys of Be known so far are metallic, which obstructs the application of Be in the fields of electric device and energy harvest where an energy gap is required.

On the other hand, it is known that by alloying Be with other elements, novel electronic structure and bizarre properties can emerge under high-pressure, including the reduction of electron gas dimensionality(*12*), superconductivity(*13, 14*), colossal valence state $Be^{8-}$ in electride(*15*), and possible LA-TA splitting(*16*). Iron (Fe) is the fourth most abundant element in nature. Its compounds are widely distributed on Earth. As early as the 1930s, researchers had conducted comprehensive experimental studies on the structure of beryllium-iron binary alloys(*17-21*). To date, three ordered Be-Fe binary phases have been reported: the ferromagnetic (FM) metallic phases of $Be_2Fe$ and $Be_5Fe$, and a still controversial non-magnetic (NM) beryllium-rich phase whose structure is undetermined. Notably, the $Be_5Fe$ phase was believed being stable only when temperature (T) > ~900 K(*19, 22*), whereas the Be-rich compound was supposed being hexagonal with a stoichiometry of 11:1(*19*). However, inconsistencies in the Be-rich phase persists up till now, and the atomic positions in it are completely unknown(*23-26*). For example, later experiments indicated that this Be-rich phase might be tetragonal $Be_{12}Fe$(*27*). Other candidates include the $Be_{15}Fe_2$, $Be_{16}Fe_3$, $Be_{17}Fe_2$, all are derived from the structure of $RhBe_{6.6}$(*23*). Currently, it is believed that $Be_{17}Fe_2$ might be the structure with the lowest enthalpy(*28*). Unfortunately, its magnetic ordering contradicts to the available experimental data(*29*), which demonstrate that the averaged atomic magnetic moment of Be-Fe alloys decreases gradually with increasing Be content, and disappears in the Be-rich phase. This apparent discrepancy, along with the fact that all experiments carried out so far are above 700 K, implies that our knowledge about Be-Fe system is highly deficient.

In order to resolve this discrepancy and to design innovative insulating alloys of Be to revolutionarily expand its application realm, in this work a thorough and unbiased first-principles structural search was performed. Insulating intermetallics $Be_4Fe$ and $Be_{11}Fe$ are predicted, as well as metallic $Be_3Fe$ and metastable $BeFe$ and $BeFe_2$. The stoichiometry of 4:1 in $Be_4Fe$ is the first ever reported in any intermetallics of Be. Previously reported compounds of $Be_{12}Fe$, $Be_{17}Fe_2$, $Be_5Fe$, and $Be_2Fe$ are also reproduced, in which we determined that $Be_{12}Fe$ and $Be_5Fe$ are metastable. Promising thermoelectric properties in $Be_{11}Fe$ are discovered, with an impressive power factor (*PF*) among all semiconductor thermoelectric materials known so far under ambient conditions.

**Methods**

The structural search was performed using evolutionary algorithm(*30*) as implemented in the USPEX code(*31*), by combining with the density functional theory (DFT) to optimize the structure and calculate the total energy. The plane-wave pseudopotential code of VASP (*32, 33*) was employed, in which the generalized gradient approximation (GGA) of Perdew-Burke-Ernzerhof (PBE)(*34*) for the exchange-correlation functional and the projector augmented-wave potentials (PAW)(*35, 36*) to describe the electron-ion interactions was adopted. The phonon dispersion was obtained by using PHONOPY(*37*), and the lattice thermal conductivities were calculated with ShengBTE(*38*). For the electronic transport properties of $Be_{11}Fe$ and $Be_4Fe$, the Boltzmann transport equation with the relaxation time approximation (RTA) was solved, for which the electron relaxation time $\tau_{nk}$ for band *n* and wavevector ***k*** is derived from the imaginary part of the electron self-energy ($\mathrm{Im}\Sigma_{nk}$) with the electron-phonon scattering has been taken into account: $\tau_{nk} = \hbar/(2\mathrm{Im}\Sigma_{nk})$. The calculation of $\Sigma_{nk}$ employed the electron-phonon coupling theory in the EPW code(*39*) and the density functional perturbation theory(*40, 41*) in the QUANTUM

ESPRESSO package(*42*). The obtained $\tau_{n\mathbf{k}}$ were then interpolated, along with the Kohn-Sham eigenvalues obtained from the DFT, onto a finer k-points grid for accurate calculation of the electronic transport properties(*43*). Further details about the calculations are provided in the Supplementary Information (SI)(*44*).

**Results**

From the calculated formation enthalpy of all searched structures, we obtained five stable structures that sit on the convex hull: $Be_{11}Fe$, $Be_{17}Fe_2$, $Be_4Fe$, $Be_3Fe$, $Be_2Fe$, and four metastable phases: $Be_{12}Fe$, $Be_5Fe$, $BeFe$ and $BeFe_2$ (Fig. 1). The dynamical stability of all these structures are corroborated by phonon spectra calculations in which no any imaginary mode can be found (Fig. S1). Our predicted structural parameters for $Be_2Fe$ and $Be_5Fe$ are in excellent agreement with the experimental data reported previously (*19*). This confirms the reliability of our theoretical method. For $Be_2Fe$, the calculated average atomic magnetic moment is 0.59 $\mu_B$, which is very close to the experimental data of 0.61 $\mu_B$(*29, 45*); the calculated lattice parameters are 4.170 Å and 6.780 Å, compared to the experimental 4.212 Å and 6.853 Å(*19*), respectively. Interestingly, the formation enthalpy of $Be_5Fe$ is 0.062 eV/atom above the convex hull, indicating that the experimentally observed phase $Be_5Fe$ could be metastable.

In the Be-rich region, a stoichiometry of 11:1 is predicted. Nevertheless, our structure has a tetragonal lattice, rather than the hexagonal assumed by Teitel and Cohen(*19*). Moreover this newly predicted phase is nonmagnetic, and is in line with what reported by Herr et al(*29*). The $Be_{12}Fe$ phase proposed by Batchelder(*27*) and the $Be_{17}Fe_2$ phase conjectured by Burr et al(*28*) are also successfully reproduced in our structure prediction. We found that $Be_{17}Fe_2$ is ferromagnetic (with an atomic magnetic moment of 0.33 $\mu_B$ for each Fe) and sits on the convex hull; whereas $Be_{12}Fe$ is 0.037 eV/atom above the convex hull, and is thus considered as metastable. The $Be_{13}Fe$ structure, that is widely observed in other Be-based binary intermetallics with a stoichiometry of 13:1, however, is deemed to be thermodynamically unstable, due to its high formation enthalpy of +0.163 eV/atom. The previously reported discrepancy in the Be-rich compound is therefore resolved when the new $Be_{11}Fe$ phase is taken into account.

Two new compounds of Be₄Fe and Be₃Fe that have not been reported before are also predicted. Both are nonmagnetic. In particular, the discovered stoichiometry of 4:1 in Be₄Fe has never been reported in any intermetallic compounds of Be before. On the other hand, our predicted orthorhombic Be₃Fe has a structure different from the previously reported phase with the same stoichiometry of Be₃M(M=Ti, Nb, Ta)(*46, 47*), which are trigonal instead.

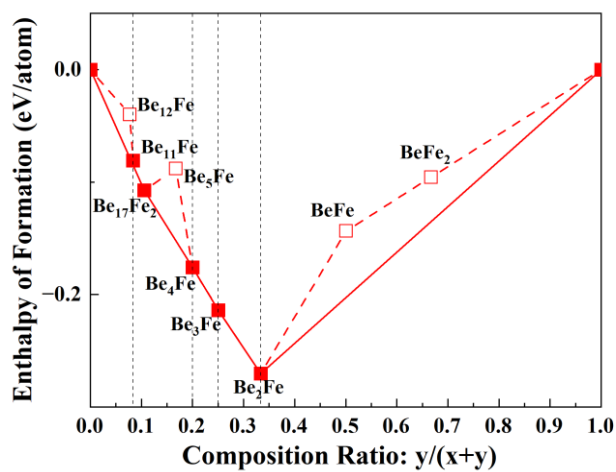

**Fig. 1. Convex hull of Be$_x$Fe$_y$.** Calculated formation enthalpy of Be$_x$Fe$_y$ with respect to the elemental solids, where the thermodynamically stable phases are represented by solid squares on the convex hull (solid line), and the metastable phases are represented by open squares connected by dashed line.

On the whole, the structure of metastable NM phases BeFe₂ and BeFe have a space group $I4/mmm$ and $P4/nmm$, respectively. As Be content increases, a series of phases Be₂Fe, Be₃Fe, Be₄Fe and metastable Be₅Fe emerge. The structure of Be₂Fe is hexagonal (space group $P6_3/mmc$), and Be₃Fe is orthorhombic and has a space group $Pmmn$, in which the Fe atoms locate at 2a sites, two Be atoms locate at 2b sites, and the remaining Be atoms locate at 4e sites, respectively [Fig. 2(a)]. The newly discovered Be₄Fe has a lower symmetry with space group $C2/c$. Its unit cell contains four Fe atoms (located at 4e sites) and sixteen Be atoms (located at two different 8f positions) [Fig. 2(b)]. The metastable Be₅Fe is cubic with a space group $F\bar{4}3m$.

The interesting Be-rich structure Be₁₁Fe is tetragonal and has space group $P\bar{4}m2$, in which the Fe atom locates at the body center (1c site) and is surrounded by a Be₁₆ cage. All Be atoms are divided into two groups: one at the vertex (1a site), and the remaining Be atoms (sitting at 2g, 4j and 4k sites, respectively) form a clathrate structure with 28 faces to enclose the central Fe atom. It is worth noting that, although not pointed out in previous reports on $Be_{17}Fe_2$(*28*) and $Be_{12}Fe$(*27*), in fact they all are clathrate structures, in which the Fe atoms are enclosed by Be cages with 28 and 34 faces, respectively [Fig. S2]. Specifically, the four neighboring Be atoms at the junction of the $Be_{16}$ cages in $Be_{11}Fe$ form an irregular tetrahedron (similar to the tetrahedron in a body centered cubic lattice). By inspecting the electron localization function (ELF) and real space charge density distribution, we found remarkable charge localizations at the center of these tetrahedrons, that form interstitial quasi-atom (ISQ) [Fig. 2(c)]. The ELF value of these ISQs reach as high as more than 0.74, far beyond that of electron gas. It is worth mentioning that these ELF maxima also correspond to the charge density maxima exactly (Fig. S4), confirming that $Be_{11}Fe$ is a clathrate electride(*15*). The charge localization in the metastable $Be_{12}Fe$ is similar, which is also a clathrate electride.

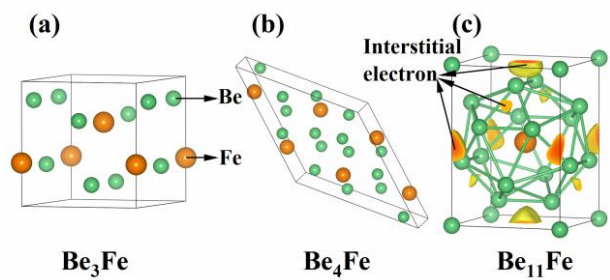

**Fig. 2. Crystal structure and ELF of newly predicted stable Be-Fe compounds.** (color online) (a) $Be_3Fe$ with space group $Pmmn$; (b) $Be_4Fe$ with space group $C2/c$; (c) $Be_{11}Fe$ with space group $P\bar{4}m2$, in which the isosurface of ELF is set to 0.74.

It is interesting to note that interstitial electron is also observed in $Be_{17}Fe_2$, $Be_5Fe$, and $Be_2Fe$, but their ELF values are too low to be considered as conventional ISQs. However, with the Bader charge analysis(*48*), we found unequivocal charge accumulation into these interstitial sites. Namely, they correspond to the centers of independent Bader volumes without any atom in it that separated by zero-flux surface of charge density distribution (Fig. S3, Table. 1). By the definition of charge accumulation and polarization, they form a kind of *de facto* ISQ. The difference from conventional ISQ is here the electrons will not dwell there perpetually: The electrons flowing into these regions will flow out latterly, thus leading to a low ELF. It is a new type of ISQ, we can call them type-II ISQs or fluxible ISQs. As listed in Table 1, both types of ISQs have significant charge amounts. In $Be_{12}Fe$, each of its eight equivalent ISQs can hold a charge of up to $2.7e^-$, which is close to the number of valence electrons of a Be atom. Most ISQs are located in the tetrahedral interstitial sites formed by Be atoms. However, all type-II ISQs in $Be_2Fe$ and some type-II ISQs in $Be_{17}Fe_2$ are uniquely located at the center of hexahedral sites. These sites are at the center of a triangle formed by three Be atoms, with one atom directly above and below this center. These two atoms are allowed to be Fe atoms, such as the 1f site in $Be_{17}Fe_2$ and the 2c site in $Be_2Fe$. When these two atoms are Fe atoms, the localized charge at these sites is significantly reduced compared to the interstitial sites fully surrounded by Be atoms, suggesting that the formation of ISQs is primarily associated with Be.

**Table 1.** Information on the interstitial sites with charge accumulation in $Be_{12}Fe$, $Be_{11}Fe$, $Be_{17}Fe_2$, $Be_5Fe$, and $Be_2Fe$, respectively.

| Phase | Type of interstitial sites | Wyckoff site | Interstitial charge($e^-$) | Interstitial coordinates (fractional) | Type of ISQ |
|---|---|---|---|---|---|
| $Be_{12}Fe$ | tetrahedral | 8h | 2.698 | 0.688 0.688 0.688 | ISQ |
| $Be_{11}Fe$ | tetrahedral | 2g | 2.557 | 0.500 0.000 0.265 | |
| | | 1d | 2.143 | 0.000 0.000 0.500 | |
| | | 1b | 2.253 | 0.500 0.500 1.000 | |
| $Be_{17}Fe_2$ | tetrahedral | 2h | 2.300 | 0.333 0.667 0.185 | type-II ISQ |
| | hexahedral | 1f | 1.407 | 0.667 0.333 0.500 | |
| | | 2g | 1.720 | 0.000 0.000 0.185 | |
| | | 1d | 2.482 | 0.333 0.667 0.500 | |
| $Be_5Fe$ | tetrahedral | 4c | 2.216 | 0.250 0.250 0.250 | |
| | | 4b | 1.882 | 0.500 0.500 0.500 | |
| $Be_2Fe$ | hexahedral | 2b | 2.087 | 0.000 0.000 0.250 | |
| | | 2c | 0.989 | 0.333 0.667 0.250 | |

In all of these compounds, a strong charge transfer between Be and Fe atoms is observed, and Fe always acts as anions. This is in contrast to the usual behavior of Fe-based intermetallics, where Fe usually acts as cations. It also defies the electronegativity argument, since Fe and Be have similar electronegativity at 0 GPa(*49*), they should not have large charge transfer if the electronegativity is valid. In particular, we found the charge loss in each Be atom keeps almost constant (about 1 $e^-$/atom) spanning the whole stoichiometries. As shown in Fig.3(b), in $BeFe_2$, $BeFe$, $Be_3Fe$, and $Be_4Fe$, all of the lost charges go to the Fe atom, making the latter attain a nominal valence state of -0.5, -1, -3, and -5, respectively. It is necessary to point out that the charge gain of Fe in $Be_4Fe$ can reach as high as 4.8 $e^-$/atom, suggesting the whole *d*-shell of the Fe atom has been fully occupied in this compound. This is the highest negative charge state of Fe atom in all of Fe-related compounds known so far, and is quite unusual. In all other compounds of $Be_2Fe$, $Be_5Fe$, $Be_{17}Fe_2$, $Be_{11}Fe$, and $Be_{12}Fe$, in addition to the charge transfer to Fe atoms, the left charges are all transferred to the interstitial sites, forming conventional ISQs or type-II ISQs.

On the other hand, the charge transfer based on Bader charge analysis can also be qualitatively verified through the electronic density of states (DOS). Take $Be_4Fe$ and $Be_3Fe$ that have high negative charge states of Fe as examples. Following the strategy used by Refs(*50, 51*), we removed all Be atoms from the original configurations of these compounds. The total density of states (TDOS) of the remaining Fe atoms is similar to that of bcc Fe (ignoring spin) (Fig. 3(c)), and there is a significant contribution in the partially unoccupied states. However, when Be atoms are inserted, the projected density of states (PDOS) of Fe shows that the Fermi level is shifted up and forming a gap or pseudogap (Fig. 4(a, c) and Fig. S6), indicating that the previously unoccupied orbitals of Fe are now filled with electrons gained from Be.

Besides $Be_4Fe$ and $Be_3Fe$, because of this strong charge transfer, the compounds of $Be_{11}Fe$, $Be_2Fe$ and metastable $BeFe$ also exhibit a narrow band gap or pseudogap at the Fermi level [Fig. 3(a)]; and $Be_{17}Fe_2$ and metastable $Be_{12}Fe$ and $BeFe_2$ display a pseudogap slightly away from the Fermi level. Detailed Bader charge and density of states (DOS) analysis suggest that Fe gains extra electrons to directly fill its own *3d* orbitals like the anions in Zintl phases, which also hybridizes with Be-*2p* orbitals.

Moreover, the mixing and splitting of the *3d* and *4p*-based bands of Fe and the hybridization of *p-d* orbitals with Be also contribute to the formation of the gap or pseudogap, similar to the mechanism of Zintl phases and the 18-n rule phases(*52-54*). This makes binary Be-Fe compounds distinctly different from conventional intermetallics that have strong metallic behavior.

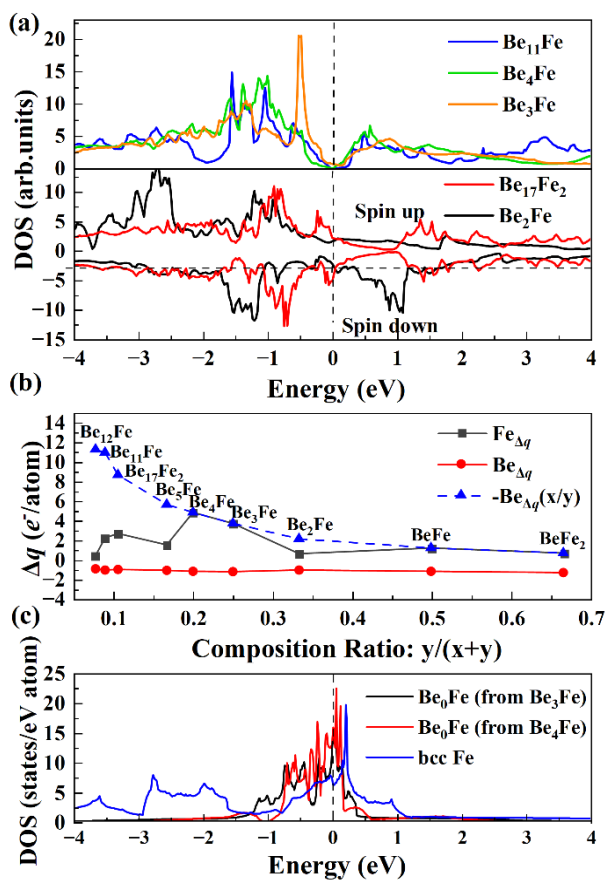

**Fig. 3. Total density of states and Bader charges.** (color online) (a) TDOS of stable Be-Fe compounds, where the Fermi level is set to 0 eV. (b) The averaged Bader charge of Be, Fe, and -Be(x/y) (the total loss of the charge of all Be atoms in the unit cell with respect to the number of Fe atoms) in $Be_xFe_y$. (c) TDOS of the remaining Fe atoms after removing Be atoms from the original configurations of $Be_3Fe$ and $Be_4Fe$, and the TDOS of bcc Fe (no spin polarization).

We now turn to the band structure of $Be_{11}Fe$ and $Be_4Fe$, both of which exhibit a narrow gap at the Fermi level. The PDOS indicate that their electronic structure near the Fermi level are predominantly governed by Fe-$3d$ and Be-$2p$ orbitals, as well as their hybridization. Figures 4(a, c) display the band structures of both compounds calculated at the PBE level, in which a small indirect band gap can be observed. For Be$_{11}$Fe, the valence band maximum (VBM) and conduction band minimum (CBM) are located at the A and Z points, respectively. While for the lower-symmetric Be$_4$Fe, the VBM and CBM do not locate at high-symmetry points.

The calculated band gap for Be$_{11}$Fe and Be$_4$Fe at the level of PBE are 0.06 eV and 0.04 eV, respectively. The more accurate gaps obtained by single-shot GW calculations (G$_0$W$_0$) are 0.22 eV and 0.85 eV, respectively. The total DOS are shown in Figs. 4(b, d). It is necessary to note that such size of a band gap is typical for many semiconductors, suggesting that Be$_4$Fe and Be$_{11}$Fe could possess prominent semiconducting properties. In fact, similar magnitude of band gap is not uncommon in Fe-based ternary Heusler compounds and Fe-Si/Ga compounds. They have high power factors (PF) due to high Seebeck coefficients and low resistivities(*55-58*). This, together with the clathrate structure, implies that Be$_{11}$Fe might be promising thermoelectric materials(*59*). Furthermore, the finite band gap and interstitial charge localization in Be$_{11}$Fe suggest that it might possess universal metallic suface states, as indicated in Ref(*60*).

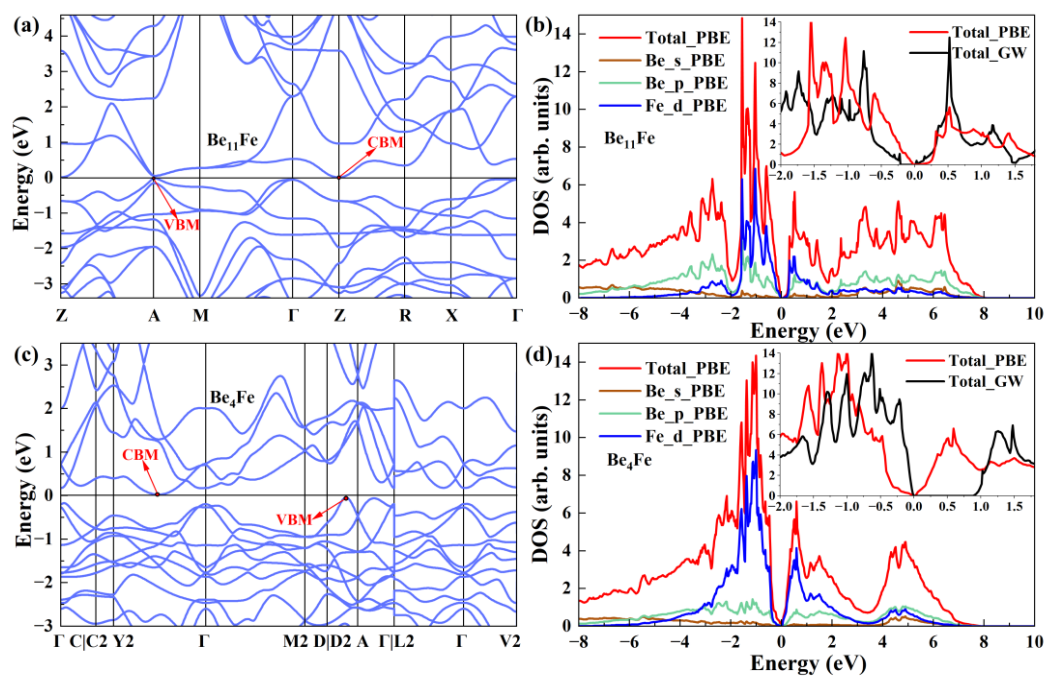

**Fig. 4. Electronic structure of Be$_4$Fe and Be$_{11}$Fe.** (color online) Electronic band structure of (a) Be$_{11}$Fe and (c) Be$_4$Fe calculated by PBE, respectively. (b, d) The projected density of states and TDOS of Be$_{11}$Fe and Be$_4$Fe calculated by PBE, with the TDOS calculated by G$_0$W$_0$ method (black) compared with those of PBE (red) in the inset. All Fermi levels are set to 0 eV.

The calculated electrical and thermal transport properties along the three cartesian directions (*x, y, z*) for Be$_{11}$Fe and Be$_4$Fe are plotted in Fig. 5, respectively. Due to the symmetry, the *x* and *y* directions of Be$_{11}$Fe have identical properties. Remarkably, Be$_{11}$Fe exhibits an extremely high intrinsic electrical conductivity [Fig. 5(a)], reaching a maximum of $4.4 \times 10^5$ $\Omega^{-1} m^{-1}$ at 400 K along the *x* and *y* directions, more than five folds higher than the typical semi-metallic Heusler compound Fe$_2$VAl(*57*). In contrast, Be$_4$Fe has a typical semiconductor conductivity behavior that increases gradually with temperature. This striking difference might be attributed to the weak electron-phonon scattering in Be$_{11}$Fe which leads to a longer electron relaxation time than in Be$_4$Fe.

To justify this argument quantitatively, we approximate the relaxation time $\tau_{nk}$ of band *n* at wave vector ***k*** by a constant tensor $\tau_c$, which is expressed as:

$$\tau_{c_{\alpha\beta}} = \frac{\sigma_{\alpha\beta}(\mu, T)V}{\sum_{nk} v_{nk\alpha} v_{nk\beta} \left[-\frac{\partial f_\mu(\epsilon_{nk}, T)}{\partial \epsilon_{nk}}\right]} \quad (1)$$

Here, $\sigma_{\alpha\beta}$ is the electrical conductivity tensor under the RTA approximation. $v_{nk\alpha}$ and $\epsilon_{nk}$ are the electron group velocity and eigen-energy of the state (*n*, ***k***), respectively; and $f_\mu(\epsilon_{nk}, T)$ is the Fermi-Dirac distribution function. $V$, $T$, and $\mu$ are the cell volume, temperature, and chemical potential, respectively. At the same temperature, our calculation shows that $\tau_c$ of Be$_{11}$Fe is 2-3 orders of magnitude larger than that of Be$_4$Fe [Fig. 5(b)]. Furthermore, due to the small bandgap of Be$_{11}$Fe, increasing temperature will make more valence electrons jump to the conduction band and leave holes behind, which enhances the carrier concentration and thus the conductivity. However, this effect diminishes with increased temperature. On the other hand, the $\tau_c$ of Be$_{11}$Fe rapidly decreases with increasing temperature (especially within 200-450 K), and impacts the conductivity negatively. It is the competition of these two opposite effects that cause the maximum in the conductivity of Be$_{11}$Fe at about 450 K, as shown in Fig. 5(a).

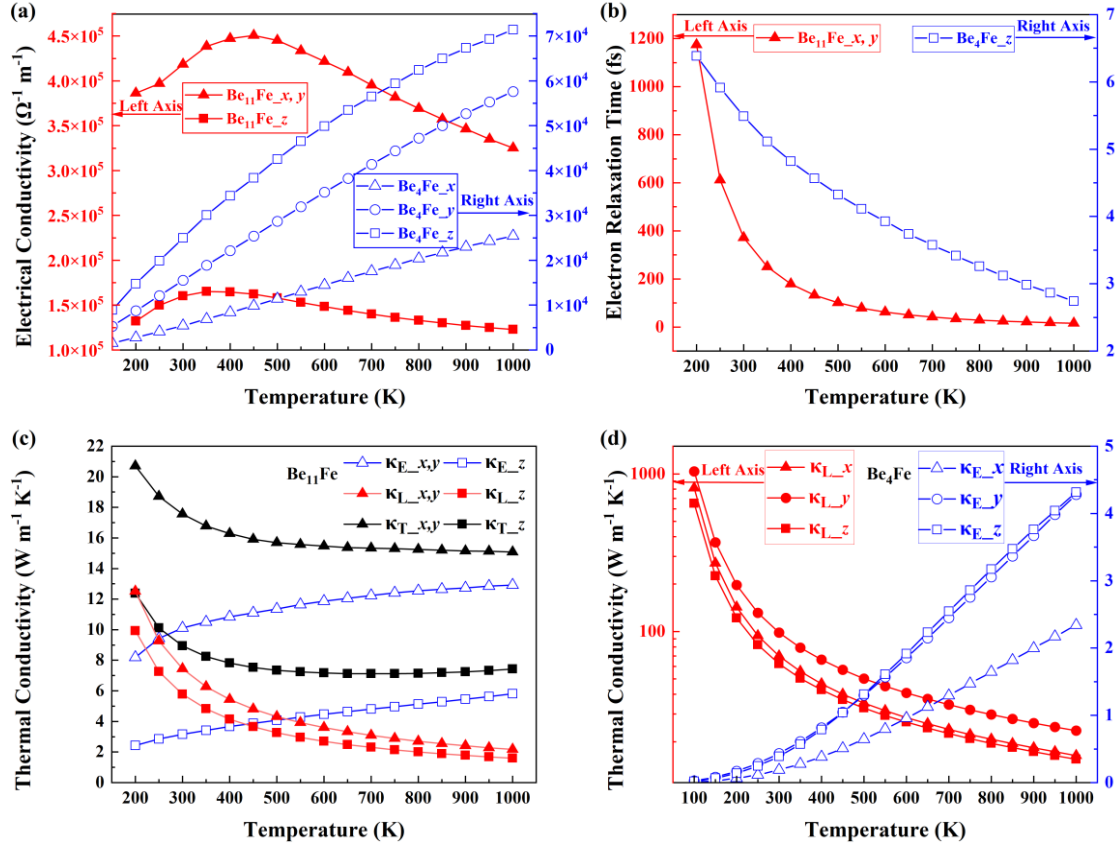

**Fig. 5. Transport properties of Be$_4$Fe and Be$_{11}$Fe.** (color online) (a) Calculated intrinsic electrical conductivity of Be$_{11}$Fe and Be$_4$Fe. (b) The calculated $\tau_c$ for along the *x* and *y* directions of Be$_{11}$Fe and along the *z* direction of Be$_4$Fe, respectively. (c) The lattice thermal conductivity ($\kappa_L$), electronic thermal conductivity ($\kappa_E$), and their sum ($\kappa_T$) of Be$_{11}$Fe, along different directions, respectively. (d) The electronic thermal conductivity (right axis) and lattice thermal conductivity (left axis) of Be$_4$Fe, respectively.

The long electron relaxation time of Be$_{11}$Fe also endows it with a considerable electronic thermal conductivity, as shown in Fig. 5(c). Its electronic thermal conductivities along the (*x, y*) and *z* directions are 10.12 and 3.17 $W\,m^{-1}K^{-1}$ at 300 K, respectively, which increase with the temperature and eventually surpass the lattice thermal conductivities, that are only 7.44 and 5.78 $W\,m^{-1}K^{-1}$ along the same respective directions at 300 K. In the range of 200-1000 K, the total thermal conductivities of Be$_{11}$Fe along the (*x, y*) directions reach about 15-20 $W\,m^{-1}K^{-1}$, which is very large if compared to typical thermoelectric materials. By comparison, the thermal conductivity of Be$_4$Fe is even larger, with a total thermal conductivity of 98.98 $W\ m^{-1}K^{-1}$ along the *y* direction at 300 K. It mainly originates from the lattice thermal conductivity, that reaches a striking 1036.18 $W\ m^{-1}K^{-1}$ at 100 K. In contrast, the maximum electronic thermal conductivity of Be$_4$Fe within the considered temperature range does not exceed 4 $W\ m^{-1}K^{-1}$, and can be neglected. Such a large lattice thermal conductivity in Be$_4$Fe implies that the scattering of phonons is very weak, thus being an ideal harmonic solid. This observation is in line with its small Grüneisen parameter of 0.3 at 300 K.

For Be$_{11}$Fe, if assuming that a moderate hole doping does not cause significant change in the electronic structure, its electrical conductivity can be further enhanced substantially. Our calculations show that if the chemical potential is shifted downward by -0.06 eV from the Fermi level at 300 K, the electrical conductivity along *x* and *y* can reach $3.5 \times 10^6$ $\Omega^{-1}\ m^{-1}$ [Fig. 6(a)]. Such a large conductivity implies that it will have a very high *PF* as long as its Seebeck coefficient is not very small. Indeed, the calculated Seebeck coefficient of Be$_{11}$Fe at 300K is shown in Fig. 6(b), in which the hole-doped Seebeck coefficient is slightly higher than the electron-doped one, and along the *x*, *y* directions it reaches a maximal value of 141.7 $\mu V\ K^{-1}$ at -0.002 eV. In this way, Be$_{11}$Fe achieves an impressive *PF* of 184 μW cm$^{-1}$ K² at 200 K, the highest among all semiconductor thermoelectric materials under ambient conditions.

In Fig. 6(d), we compare the *PF* of Be$_{11}$Fe with typical semiconductor thermoelectric materials. Be$_{11}$Fe possesses a significant advantage, with an optimal *PF* reaches 178 $\mu W\ cm^{-1}\ K^2$ at 300K (chemical potential shift -0.004 eV), much higher than the 106 $\mu W\ cm^{-1}\ K^2$ of the half-Hessler compound Nb$_{0.95}$Ti$_{0.05}$FeSb at the same temperature(*61*). Furthermore, at the chemical potential that optimizes the *PF*, Be$_{11}$Fe has a respectable *ZT* value of 0.36 [Fig. 6(c)], and with a large room for further improvement. At 300 K, the lattice thermal conductivity of Be$_{11}$Fe is close to 9 $W\ m^{-1}K^{-1}$, indicating that it can be reduced further. Given its ultra-high *PF*, reducing the lattice thermal conductivity would yield substantial benefits. Methods such as doping, alloying, and nanostructuring are expected to significantly increase its *ZT* value by reducing the lattice thermal conductivity. Additionally, adjusting the band gap of Be$_{11}$Fe and utilizing the universal metallic surface states of electride(*60*) to achieve a better balance between the Seebeck coefficient, electrical conductivity, and electronic thermal conductivity is also another feasible approach to enhance its *ZT* while maintaining a high *PF*.

The properties of Be$_{11}$Fe, characterized by its ultra-high *PF* and relatively respectable *ZT*, make it highly suitable for waste heat energy harvest applications that with large heat source capacity or low heat source cost(*61, 62*), as well as active cooling fields that require high *PF* and high thermal conductivity rather than high *ZT(63-65)*. Furthermore, Be-rich compounds generally possess the advantages of being lightweight, high strength, and highly oxidation-resistant(*66, 67*), which further enhances the practical application prospects of Be$_{11}$Fe. In contrast, due to the small electrical conductivity and large thermal conductivity, Be$_4$Fe is not suitable when applied as thermoelectric. Nonetheless, according to the Shockley–Queisser model(*68*), its band gap of 0.85eV implies that it might have good photovoltaic property.

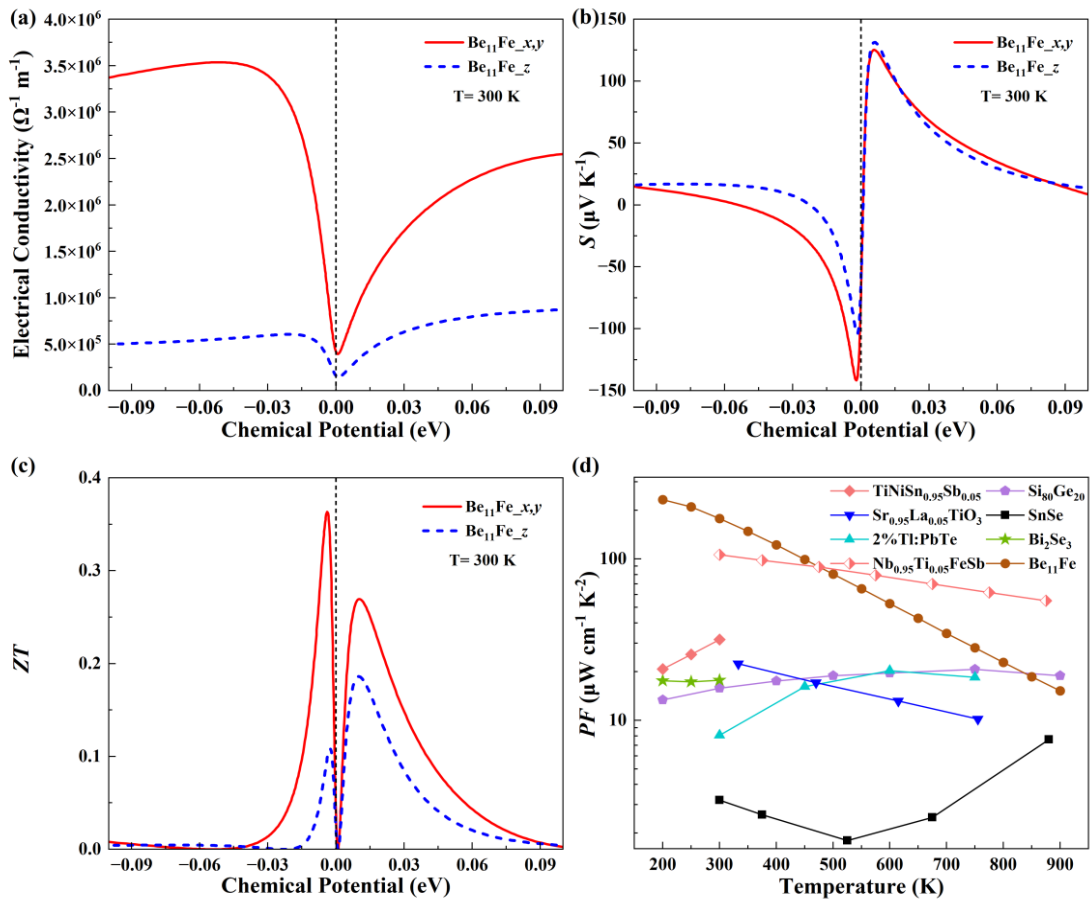

**Fig. 6. Thermal electric properties of Be$_{11}$Fe**. (color online) (a) The electrical conductivity, (b)

Seebeck coefficient, and (c) figure of merit *ZT* of Be$_{11}$Fe as a function of chemical potential at 300 K, respectively. (d) Comparison of the *PF* of Be$_{11}$Fe in (*x*, *y*) directions to the optimal *PF* of typical bulk thermoelectric materials reported previously (including TiNiSn(*69*), Si$_{20}$Ge$_{80}$(*70*), SnSe(*71*), PbTe(*72*), SrTiO$_3$(*73*), Bi$_2$Se$_3$(*74*), Nb$_{0.95}$Ti$_{0.05}$FeSb(*61*)).

## Conclusion

In summary, five new structures in Be-Fe intermetallics unknown so far were discovered. The longstanding discrepancy in Be-rich compound was resolved with the new Be$_{11}$Fe phase, which is a clathrate structure with the central Fe atoms are surrounded by a 28-face cage formed by Be atoms. We found that in all Be-Fe intermetallics, Fe acts as an anion, against the usual wisdom and the argument based on electronegativity. In particular, the valence state of Fe can reach -5, the highest negative charge state known for all Fe-based compounds. Moreover, we identified a new type of ISQ in Be$_{17}$Fe$_2$, Be$_5$Fe and Be$_2$Fe, which accumulate excess charge in interstitial sites but with a low ELF. We also found that a little bit of Fe does a lot for Be, leading to insulating phases (Be$_{11}$Fe and Be$_4$Fe) that are the first in all known Be-based intermetallic compounds. We also discovered that Be$_{11}$Fe not only has the highest stoichiometry in Be-Fe binary, but has an ultra-high *PF* value (178 $\mu W\ cm^{-1}\ K^2$ at 300K), and a respectable *ZT* value (0.36 at 300K), making it a promising material in waste heat energy harvest and active cooling. Furthermore, the lightweight nature of Be$_{11}$Fe indicates its potential to serve as a backup energy system for planetary probes. These findings not only deepen our understanding about the Be-based compounds, but also expand the application realm of Be from solely being used as supportive mechanical component to wide functional areas such as thermoelectrics, photovoltaics, and electronic devices, etc., that has superior multifunctional properties simultaneously.

## Author Contribution

## Conflicts of interest

There are no conflicts to declare.

## Acknowledgments

This work was supported by National Key R&D Program of China under Grant No. 2021YFB3802300, the NSAF under Grant Nos. U1730248, the National Natural Science Foundation of China under Grant No. 12074274, and the Foundation of National key Laboratory of shock wave and Detonation physics under Grant No. 2023JCJQLB05401. Part of the calculations were performed using the resources provided by the center for Comput. Mater. Sci. (CCMS) at Tohoku University, Japan.

**Data availability**

The data that support the findings of this study are available from the corresponding author upon reasonable request.

# Supplementary Information For
# Distinctive Electronic Characteristics and Ultra-high Thermoelectric Power Factor in Be-Fe Intermetallics

Qi D. Hao[1, 2], H. Wang[2], Xiang R. Chen[1]*, Hua Y. Geng[2, 3]*

[1] *College of Physics, Sichuan University, Chengdu 610065, P. R. China;*

[2] *National Key Laboratory of Shock Wave and Detonation Physics, Institute of Fluid Physics, China Academy of Engineering Physics, Mianyang, Sichuan 621900, P. R. China;*

3 *HEDPS, Center for Applied Physics and Technology, and College of Engineering, Peking University, Beijing 100871, P. R. China;*

* *Corresponding authors. E-mail: s102genghy@caep.cn; xrchen@scu.edu.cn*

## Supplementary Methods

We performed two independent variable-composition structure searches using USPEX code[1] to find possible stable phases in Be-Fe system, covering a size of the system up to 28 atoms/cell. The maximal generation was set to 60. For each structure that might be stable, we performed higher precision calculations using VASP[2,3] again, where the employed structural relaxation criterion is that for the forces on each atom must be less than 0.001 eV/Å. The electronic SCF criterion was set to a tolerance of $10^{-7}$ eV, the plane wave basis cut-off energy was set to 700 eV, and Monkhorst-Pack[4] k-meshes with a grid spacing of $2\pi \times 0.016$ Å$^{-1}$ that including the Γ point were used. The formation enthalpy per atom of $Be_xFe_y$ was calculated by $\Delta H(Be_xFe_y)=[H(Be_xFe_y)-xH(Be)-yH(Fe)]/(x+y)$, where H is the total enthalpy. We also note that $Be_{11}Fe$ has an enthalpy less than 4 meV/atom higher than the convex hull, which is too small to be distinguished with the accuracy of the available theoretical methods. We thus take it as a ground state in this work as well.

$G_0W_0$ calculations were performed on top of the PBE wavefunctions, where the energy cutoff for the response function was set to 350 eV. The plane wave cutoff energy (ENCUT) was set to 650 eV, and the maximum number of bands supported by this value was used for exact diagonalization of the Kohn-Sham Hamiltonian. For $Be_{11}Fe$ and $Be_4Fe$, the Monkhorst-Pack k-meshes of 9×9×7 and 8×8×6 including the Γ point were used respectively. The Bader charge analysis data were calculated using a fine Fast Fourier Transform grid of $12 \times G_{cut}$ ($G_{cut} = \frac{\sqrt{2m_e ENCUT}}{\hbar}$), which is sufficient to ensure the convergence of the Bader charge analysis for all phases. We calculated the lattice thermal conductivity using the ShengBTE code[5] by considering the influence of third-nearest neighbor atoms, where the required second- and third-order force constants were obtained using a 2×2×2 supercell. In DFPT calculations with QUANTUM ESPRESSO package[6], norm-conserving pseudopotentials were used, and as in VASP calculations, $1s^22s^2$ of Be and $3s^23p^64s^23d^6$ of Fe were treated as valence electrons. The kinetic energy cutoff for wavefunction expansion was set to 93 Ry, and q-points meshes of 6×6×6 and 4×4×4 were used for $Be_{11}Fe$ and $Be_4Fe$, respectively. Although such accuracy is enough to obtain sufficiently accurate phonon frequencies, it is not fine enough for calculating the electron self-energy through the electron-phonon coupling theory. We thus further performed interpolation using EPW code[7] and finally executed the calculations on a 30×30×30 grid and obtained an imaginary part of the electron self-energy on a 12×12×12 k-points mesh. Then we performed a high-accuracy non-self-consistent calculation with VASP using a Monkhorst-Pack k-mesh with a grid spacing of $2\pi \times 0.011$ Å$^{-1}$ to obtain the Kohn-Sham eigenvalues. Thanks to the Fourier interpolation module provided by the BoltzTraP2 program[8], We were able to conveniently interpolate the relaxation time and the Kohn-Sham eigenvalues together to a grid that has 45 times more k-points than the one used for calculating the Kohn-Sham eigenvalues, and the accurate electronic transport properties were then calculated.

## Supplementary Figures

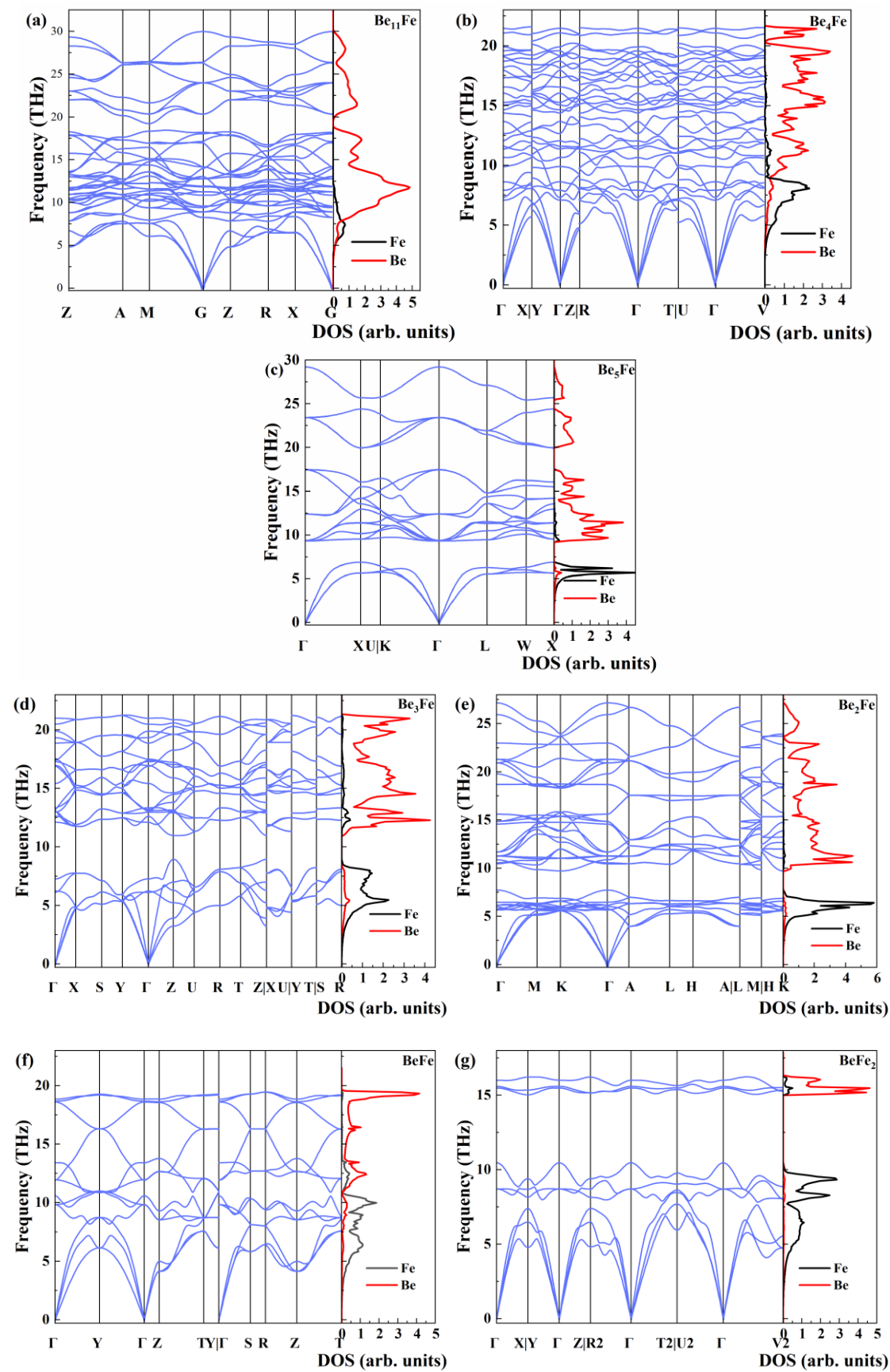

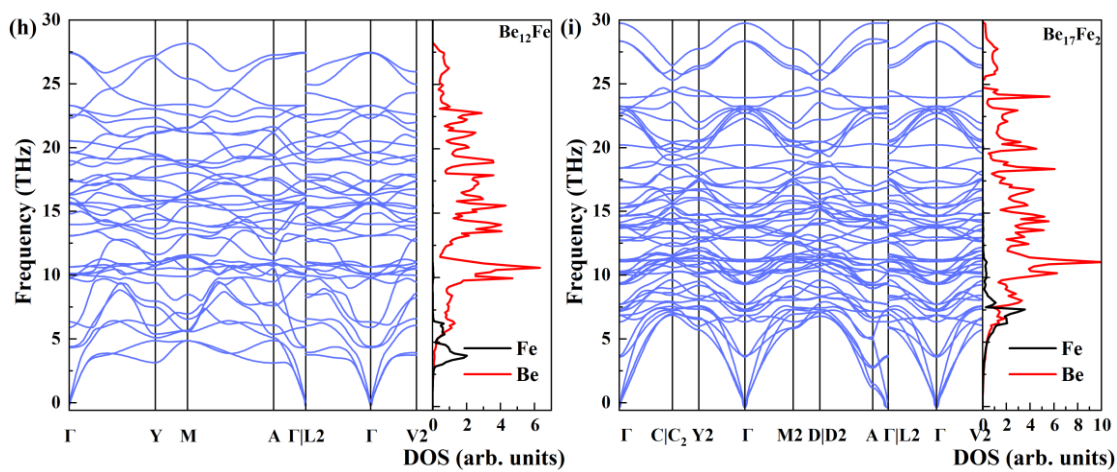

**Fig. S1.** (color online) Calculated phonon dispersion and projected phonon density of states of (a) Be$_{11}$Fe, (b) Be$_4$Fe, (c) Be$_5$Fe, (d) Be$_3$Fe, (e) Be$_2$Fe, (f) BeFe, (g) BeFe$_2$, (h) Be$_{12}$Fe, and (i) Be$_{17}$Fe$_2$, respectively.

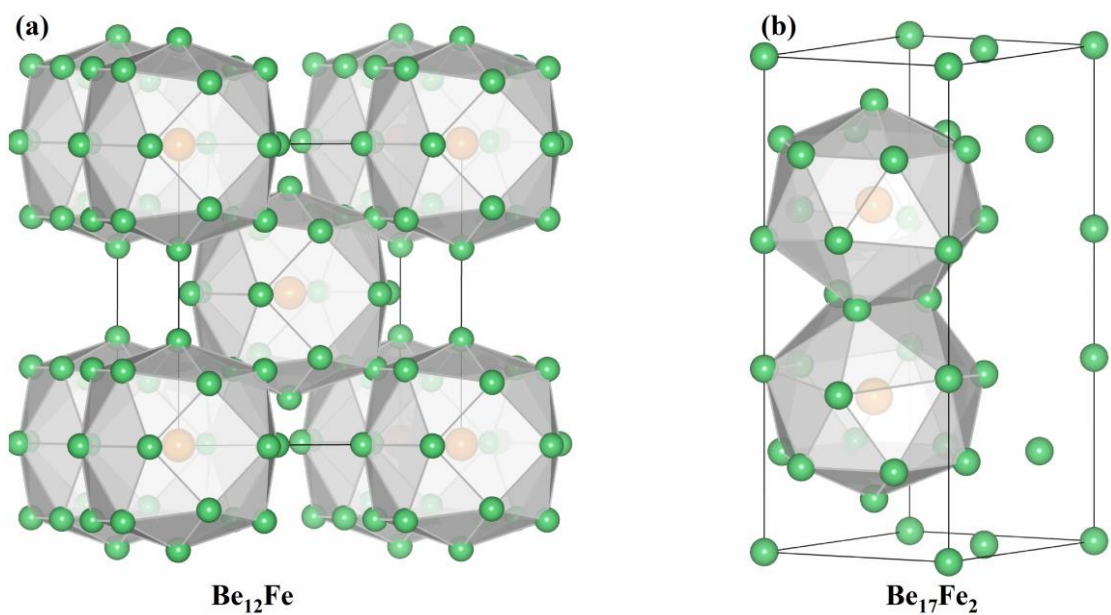

**Fig. S2.** (color online) The clathrate structures in Be$_{12}$Fe and Be$_{17}$Fe$_2$, where the Fe atoms in Be$_{12}$Fe are surrounded by 34-face cages formed by 20 Be atoms (a), and the Fe atoms in Be$_{17}$Fe$_2$ are surrounded by 28-face cages formed by 16 Be atoms (b).

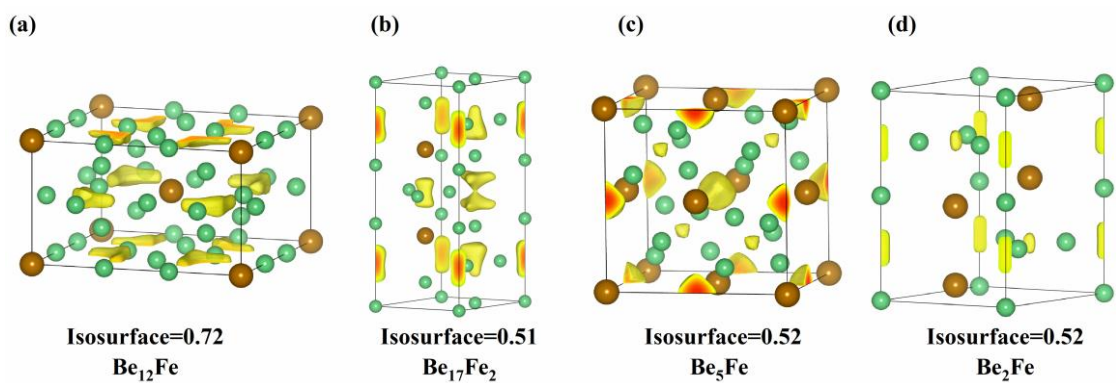

**Fig. S3.** (color online) The ELF of selected Be-Fe compounds: (a) Be$_{12}$Fe, (b) Be$_{17}$Fe$_2$, (c) Be$_5$Fe,

and (d) $Be_2Fe$.

In Figs. S4 and S5, we show the characteristics of conventional ISQ and type-II ISQ using $Be_{11}Fe$ and $Be_2Fe$ as the example, respectively. The locations of ISQs in both $Be_{11}Fe$ and $Be_2Fe$ are the joint maxima of charge density and ELF in real space. Therefore they both are electride according the "ELF+CHG" criterion. However, the difference is that the ELF value at the location of conventional ISQ is relatively high (> 0.6), while the ELF of type-II ISQ is at a level of ~0.5 and does not exceed 0.6. The low ELF of type-II ISQ indicates the electrons accumulated in these interstitial regions will flow in and out continually.

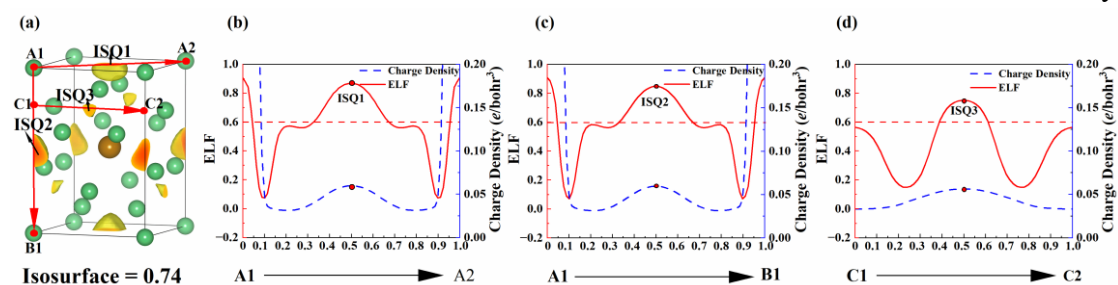

**Fig. S4.** (color online) Variation of the ELF and charge density in $Be_{11}Fe$ along the given paths. The horizontal coordinates in (b), (c) and (d) correspond to the paths from atom A1 to atom A2, from atom A1 to atom B1, and from point C1 to point C2, as shown by the arrows in (a), respectively. The midpoint of these three paths are the location of interstitial charge localization.

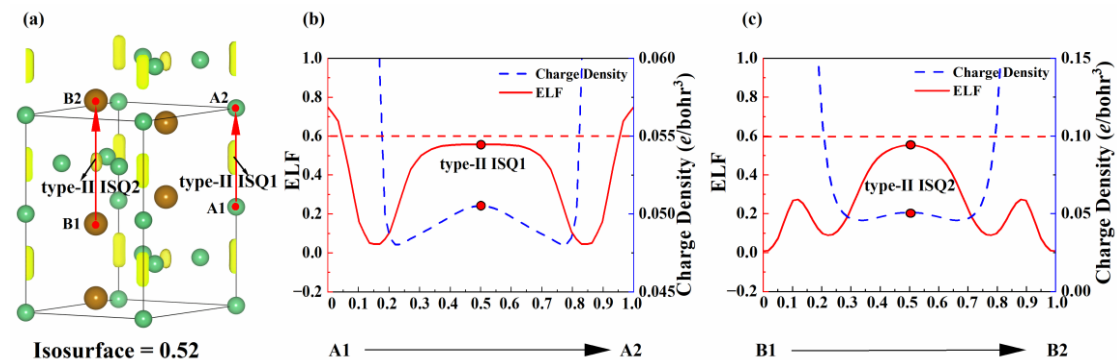

**Fig. S5.** (color online) Variation of the ELF and charge density in $Be_2Fe$ along the given paths. The horizontal coordinates in (b) and (c) correspond to the paths from atom A1 to atom A2 and from atom B1 to atom B2 as depicted by arrows in (a), respectively. The midpoint of both paths are the location of interstitial charge accumulation.

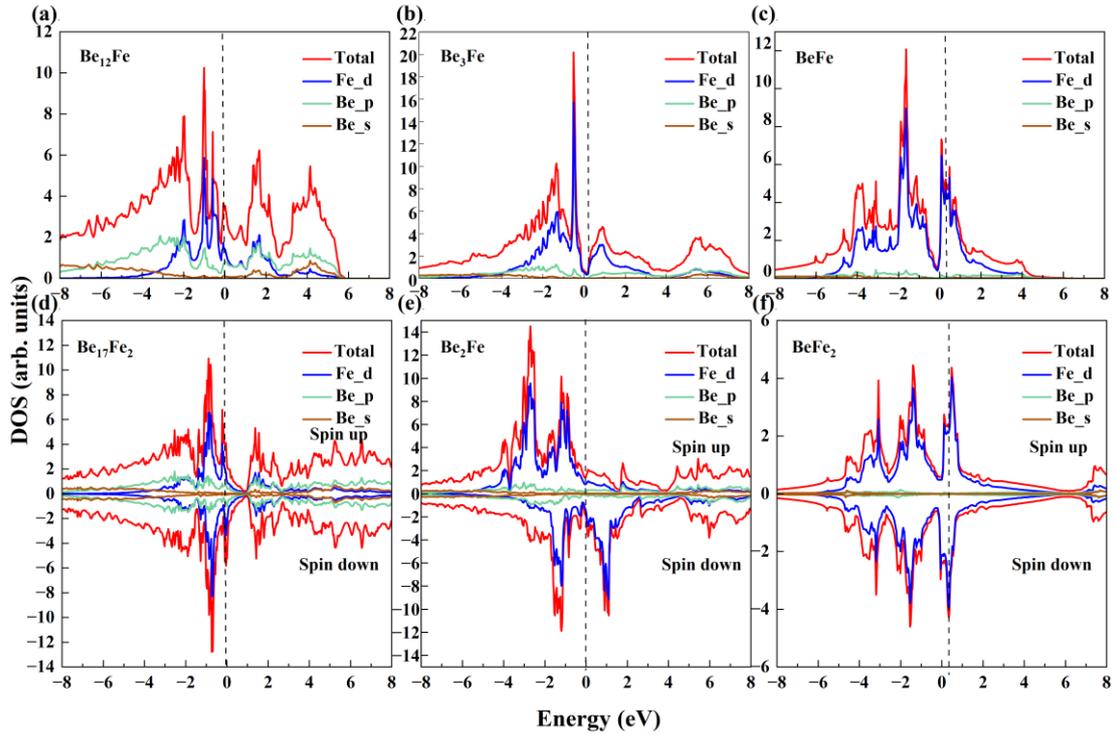

**Fig. S6.** (color online) PDOS of (a) Be$_{12}$Fe, (b) Be$_3$Fe, (c) BeFe, (d) Be$_{17}$Fe$_2$ (FM), (e) Be$_2$Fe (FM), and (f) BeFe$_2$ (FM), respectively.

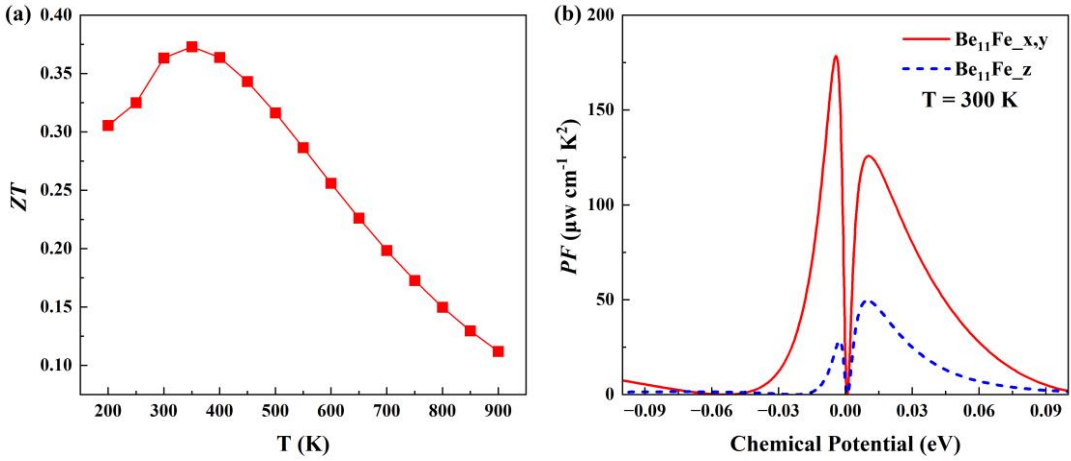

**Fig. S7.** (color online) (a) The optimal value of the figure of merit *ZT* obtained by mild hole doping at different temperatures for Be$_{11}$Fe, (b) *PF* of Be$_{11}$Fe as a function of chemical potential at 300 K.

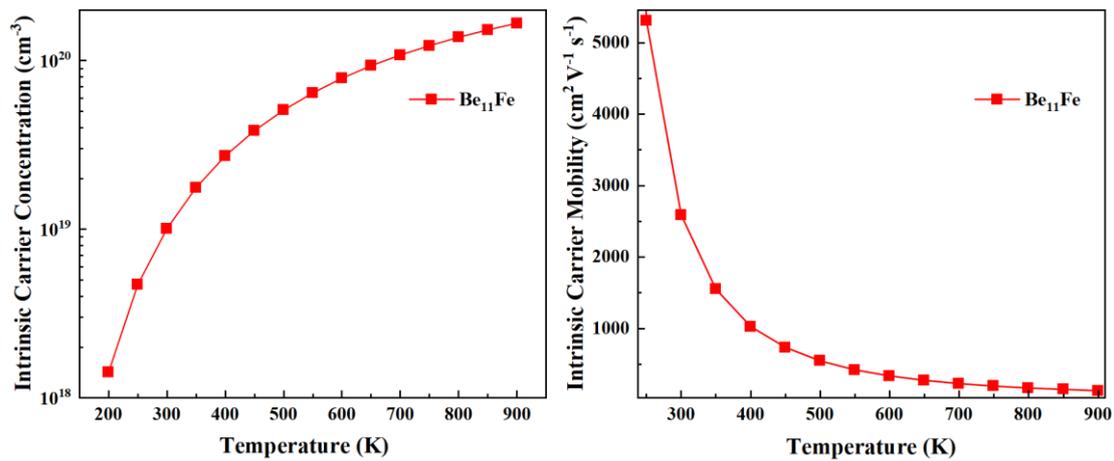

**Fig. S8.** (color online) Intrinsic carrier concentration (left) and intrinsic carrier mobility (right) of Be$_{11}$Fe, respectively.

## Supplementary Tables

Table S1. Structural information of the stable and metastable Be-Fe compounds

| Phase | Lattice parameters (Å) | Atom | Wyckoff site | Bader charge (e⁻) | Atomic coordinates (fractional) |
|---|---|---|---|---|---|
| $Be_{12}Fe$- $I4/mmm$ | a = b = 7.145<br>c = 4.083<br>α = β = γ = 90.000 | Be | 8f<br>8i<br>8j | -0.919<br>-1.025<br>-0.865 | 0.250 0.250 0.250<br>0.348 0.000 0.000<br>0.290 0.500 0.000 |
| | | Fe | 2a | 0.420 | 0.000 0.000 0.000 |
| $Be_{17}Fe_2$- $P\bar{6}m2$ | a = b = 4.090<br>c = 10.628<br>α = β = 90.000<br>γ = 120.000 | Be | 2i<br>1a<br>1c<br>2g<br>2h<br>3k<br>6n | -0.960<br>-0.784<br>-0.643<br>-1.182<br>-1.165<br>-1.127<br>-0.982 | 0.667 0.333 0.100<br>1.000 0.000 0.000<br>0.333 0.667 0.000<br>0.000 1.000 0.370<br>0.333 0.667 0.344<br>0.990 0.494 0.500<br>0.166 0.834 0.185 |
| | | Fe | 2i | 2.716 | 0.667 0.333 0.307 |
| $Be_{11}Fe$- $P\bar{4}m2$ | a = b = 4.108<br>c = 5.785<br>α = β = γ = 90.000 | Be | 1a<br>2g<br>4j<br>4k | -0.582<br>-1.146<br>-1.138<br>-1.078 | 0.000 1.000 1.000<br>1.000 0.500 0.256<br>0.744 0.000 0.382<br>0.500 0.255 0.133 |
| | | Fe | 1c | 2.217 | 0.500 0.500 0.500 |
| $Be_5Fe$- $F\bar{4}3m$ | a = b = c = 5.825<br>α = β = γ = 90.000 | Be | 4c<br>16e | -1.107<br>-1.125 | 0.250 0.250 0.250<br>0.374 0.126 0.874 |
| | | Fe | 4a | 1.513 | 0.000 0.000 0.000 |
| $Be_4Fe$- $C2/c$ | a = 7.716 b = 4.101<br>c = 7.690<br>α = γ = 90.000<br>β = 137.384 | Be | 8f<br>8f | -1.188<br>-1.231 | -0.308 -0.074 0.145<br>0.605 0.395 0.953 |
| | | Fe | 4e | 4.84 | 0.500 -0.128 0.750 |
| $Be_3Fe$- $Pmmn$ | a = 4.551<br>b = 3.647<br>c = 3.777<br>α = β = γ = 90.000 | Be | 2b<br>4f | -1.290<br>-1.199 | 1.000 0.500 0.690<br>0.252 0.500 0.163 |
| | | Fe | 2a | 3.686 | 0.500 0.500 0.653 |
| $Be_2Fe$- $P6_3/mmc$ | a = b = 4.177<br>c = 6.692<br>α = β = 90.000<br>γ = 120.000 | Be | 2a<br>6h | -1.004<br>-1.081 | 0.000 0.000 0.000<br>0.658 0.829 0.250 |
| | | Fe | 4f | 0.586 | 0.667 0.333 0.062 |
| $BeFe$- $P4/nmm$ | a = b = 2.569<br>c = 5.368<br>α = β = γ = 90.000 | Be | 2c | -1.190 | -0.000 0.500 0.109 |
| | | Fe | 2c | 1.190 | -0.000 0.500 0.636 |
| $BeFe_2$- $I4/mmm$ | a = b = 2.567<br>c = 8.407<br>α = β = γ = 90.000 | Be | 2a | -1.337 | 0.000 1.000 1.000 |
| | | Fe | 4e | 0.668 | 0.000 1.000 0.338 |

**Supplementary references**